 \documentclass[final,3p,times]{elsarticle}
\usepackage[utf8]{inputenc}
\usepackage{latexsym,amssymb,amsmath,amsfonts,amsthm,exscale}
\usepackage{graphicx, graphics}
\usepackage{algorithmic}
\usepackage[norelsize,english,ruled,lined,commentsnumbered]{algorithm2e}
\usepackage{cite}
\usepackage{xcolor}
\usepackage{aeguill}
\usepackage{psfrag}
\usepackage{multirow}
\usepackage{subfigure}
\usepackage{wrapfig}
\usepackage{setspace}
\usepackage{mathrsfs}

 \usepackage{graphicx}
 \usepackage{tikz}
 \usetikzlibrary{shapes,snakes,arrows,patterns,trees}
 \usetikzlibrary[mindmap]
 \usepackage{subfigure}
 \usepackage{xcolor}
\newdefinition{rem}{Remark}
\newproof{pf}{Proof}
\newproof{pot}{Proof of Theorem \ref{thm2}}
\usepackage{amssymb}
\usepackage{amsthm}
\usepackage{amsmath}
\usepackage{lineno}

\journal{Journal of Computational Physics}

\begin{document}

\begin{frontmatter}



\title{Parareal in time 3D numerical solver for the LWR Benchmark neutron diffusion transient model}
%\author{}
%\address{}
%---------------------------------------------------------------------------------
% \tnotetext[label1]{here label1}
%--------------------------------------------------------------------------------- 

 \author[add2,add5]{ Anne-Marie Baudron\corref{cor2}\corref{cor5}}
 \ead{anne-marie.baudron@cea.fr}

  									%:::::::::::::::::::::::::::::::

 \author[add2,add5]{Jean-Jacques Lautard\corref{cor2}\corref{cor5}}
 \ead{jean-jacques.lautard@cea.fr}
  									%:::::::::::::::::::::::::::::::
\author[add1,add2,add3]{Yvon Maday\corref{cor1}\corref{cor2}\corref{cor3}}
 \ead{maday@ann.jussieu.fr} \ead[url]{http://www.ljll.math.upmc.fr/maday}
 									%:::::::::::::::::::::::::::::::
									 \author[add2,add6]{Mohamed Kamel Riahi\corref{cor1}\corref{cor2}}
 \ead{riahi@ann.jussieu.fr} \ead[url]{http://www.ljll.math.upmc.fr/riahi}
%\fntext[label2]{ici footnote text}
 									%:::::::::::::::::::::::::::::::

 \author[add4]{Julien Salomon\corref{cor3}}
 \ead{salomon@ceremade.dauphine.fr}  \ead[url]{http://www.ceremade.dauphine.fr/~salomon}

 									%:::::::::::::::::::::::::::::::

% \cortext
 \address[add1]{Sorbonne Universit\'es, UPMC Univ Paris 06, UMR 7598, Laboratoire Jacques-Louis Lions and Institut Universitaire de France, F-75005, Paris, France.}
 %\cortext
 \address[add2]{Laboratoire de Recherche Conventionn\'e MANON, CEA/DEN/DANS/DM2S and UPMC-CNRS/LJLL.} 
 %\cortext
  \address[add3]{Brown  Univ, Division of Applied Maths, Providence, RI, USA}
% \cortext
  \address[add4]{CEREMADE, Univ Paris-Dauphine, Pl. du Mal. de Lattre de Tassigny, F-75016, Paris France.}
% \cortext
  \address[add5]{CEA-DRN/DMT/SERMA, CEN-Saclay, 91191 Gif sur Yvette Cedex, France.}
% \cortext
  \address[add6]{CMAP, Inria-Saclay and X-Ecole Polytechnique,
Route de Saclay, 
91128 Palaiseau Cedex
FRANCE.}									
%---------------------------------------------------------------------------------

\begin{abstract}
We present a parareal in time algorithm for the simulation of neutron diffusion transient
model. The method is made efficient by means of a coarse solver defined with large
time steps and steady control rods model. Using finite element for the space discretization, our implementation provides a good scalability of the algorithm. Numerical results show the efficiency of the parareal method
on large light water reactor transient model corresponding to the Langenbuch-Maurer-Werner (LMW) benchmark~\citep{LMWref}.
% We propose in this paper a parareal in time algorithm for the resolution of neutron diffusion transient model. The parallelization in time is efficient thanks to a coarse solver which is based on a stable coarse time step with a steady control rods model. The finite element implementation provides a good scalability of the algorithm, where the time discretization is based on a Euler implicit scheme. Numerical results show the efficiency of the parareal method on large light water reactor transient model corresponding to the Langenbuch-Maurer-Werner (LMW) benchmark.

%
%The accurate knowledge of the time-dependent spatial flux distribution in nuclear reactor is required for nuclear safety and design. The motivation behind the development of parallel methods for solving the energy-, space-, and time-dependent kinetic equations is not only the challenge of developing a method for solving a large set of coupled partial differential equations, but also a real need to predict the performance and assess the safety of large commercial reactors, both these presently operating and those being designed for the future. Numerical results show the efficiency of the parareal method on large light water reactor transient model.
\end{abstract}


\end{frontmatter}
% \linenumbers
\section{Introduction}
%------------------------------------------------------------------------------------------
%                                                                                  The introduction
%------------------------------------------------------------------------------------------
The accurate knowledge of the time-dependent spatial flux distribution in nuclear reactor is required for nuclear safety and design. The motivation behind the development of methods for solving the energy-, space-, and time-dependent kinetic equations is not only the challenge of developing a method for solving a large set of coupled partial differential equations, but also a real need to predict the performance and assess the safety of large commercial reactors, both these presently operating and those being designed for the future.
	
	Modern reactor core designs and safety, nowadays, depend heavily on the simulation of the reactor core and plants dynamics as well as their mutual interaction. Significant progress has been made, during the last fifteen years, in developing accurate techniques to simulate the computationally expensive reactor core models.
 Modeling the reactor core involves a large set of coupled time-dependent partial differential equations (PDEs), where the  exact model  kinetic transport equation is simplified to multi-group diffusion approximation. 
	The time-dependent multi-group neutron diffusion equation is usually used to study the transient behavior of the neutron flux  distribution on the reactor core, where the prediction of the dependencies of neutrons flux comprising dynamics in reactor core at a forward long time, is very important for nuclear safety. These diffusion equations are coupled with the dynamic of some {\it delayed neutrons}, which are called precursors.  
		  		  
	Due to a lack of availability of read-write memory in the sequential computers, it is relevant to often propose parallel methods, which solve these large scaled system, with massively parallel computers.
         Many successful works have been done in the parallelization of the neutron model's simulation. For instance~\citep{MR2674322} studies the static case i.e. eigenvalue problems with space domain decomposition methods, and a very nice strategy~\citep{Dahmani2001805},~\citep{stevephd} uses quasi-stationnary approach to accelerate the simulation.
         
% 	 We follow in this paper the physical model presented in~\citep{Dahmani2001805} where a sequential quasi-static method is presented.      
          In this paper we investigate the application of the parareal in time algorithm~\citep{ MR1842465,MR2235771} on neutron diffusion equation that governs the time-dependent flux distribution in the reactor core. The parareal in time algorithm is an iterative scheme, that enables to improve computational time with parallel simulation. In several cases, parareal in time algorithm gives an impressive rate of convergence for the linear diffusion equations or with the same efficiency with non-linearity~\citep{MR1962689}. This algorithm is studied and shows stability and convergence results~\citep{MR2235769, MR2427859, MR2334113, MR2306258, MR2235772} particularly for diffusion system and others. Also it presents efficiency in parallel computer simulation. We find a variety~\citep{MR2436108,MR2241303,MR2436107, MR2427859, MR2349453, MR2334115, MR2680929, MR2521690} of versions of this scheme that adapt~\citep{MR2401052, MR2359676, MR2334151, MR2235770, MR2765386, PhysRevE.66.057701} the original algorithm to tackle new settings. Furthermore the parareal algorithm can be easily coupled with other iterative schemes such as domain decomposition methods e.g. basic Schwarz algorithm or more complex one~\citep{099rim}, and optimal control based descent algorithm~\citep{MR1931522, MR2361898, MR2639234}. 
          
          The paper is organized as follows: After this introduction, we present at the second section the model of the kinetic of neutron inside the reactor core. The third section is devoted to the brief introduction of the parareal in time algorithm. Numerical tools are present in the fourth section where we explain how paraeal algorithm is adapted to solve the problem. We finally, in section five, present and discuss the numerical experiments that show the speedup following the fully-parallel implementation of the parareal algorithm in a parallel architecture. 

\section{Model}\label{sec:model}
%------------------------------------------------------------------------------------------
%                                                                                  The model 
%------------------------------------------------------------------------------------------

The neutron dynamics in a reactor core is governed by the kinetic transport Boltzmann's equation~\citep{MR792484}. The solution to these equations is denoted as $\Psi$ and represents the directional neutron flux, that is a function of time,			
%The directional neutron flux $\Psi$ solution~\citep[chap  XXI, section 2, Thm 3]{MR792484} of kinetic equation is a function who depends on 
of the position $\vec r$ within the reactor core $\Omega\subset\mathbb R^3$, of the velocity of neutron $\vec{V}=\sqrt{2E/m}\vec{d}$, where $\vec d$ is a unit vector that stands for the direction of the velocity,  $E$ stands for the energy of the neutron and $m$ for it's mass. For computational reason, a simplification of the model has been proposed in~\citep[chap XXI, sec 5]{MR792484} that consists in averaging through the velocity directional variable leading to the introduction of the new function $\phi(t,\vec r,E)=\frac{1}{4\pi}\int_{S}\Psi(t,\vec r,\vec d,E) \ d \vec{d}$ where $S$ is the unit sphere. This method leads to accurate enough results in most of standard cases, unfortunately the computational time remains excessively long. Further simplifications consist in averaging also in the energy variable : the energy interval $[E_\text{min},E_\text{max}]$ is divided into $\hat g$ non overlapping intervals around a set of discreet energies $\{E^{g}\}_{g=1}^{\hat g}$ and leads to consider a new unknown $\Phi=\{\phi^{g}\}_{g=1}^{\hat g}$ composed of the set of the neutron average flux  over each subinterval  around $E^{g}$.
 This approach is known as the multi-group theory~\citep{Reuss:1087175} where for each energy group $g=1,\hdots,\hat g$ and any position $\vec{r}\in{\Omega}\subset\mathbb R^3$, the equations are a set of coupled-three-dimensional, multi-energy-group neutron kinetics equations involving a time delayed contributions (called precursors, and denoted as $C\equiv C(\vec r,t) = \{C^{k}(\vec r,t)\}_{k=1}^K$).	 The partial differential equation that governs the kinetic of neutron in the reactor core writes
 \begin{equation}\label{eqBLZM}
\left\{\begin{array}{ll}
\dfrac{1}{V^{g}}\dfrac{\partial}{\partial {t}}\phi^{g}(\vec r,t)=&\text{div}(D^{g}\vec{\nabla}\phi^{g}(\vec r,t))-\Sigma_{t}^{g}\phi^{g}(\vec r,t)+\chi^{g}_p\displaystyle\sum_{g'=1}^{\hat g}(1-\beta^{g'})\nu^{g'}\Sigma_{f}^{g'}\phi^{g'}(\vec r,t)\\
&+\displaystyle\sum_{g'=1}^{\hat g}\Sigma_{s}^{g'\rightharpoonup g}\phi^{g'}(\vec r,t)+\displaystyle\sum_{k=1}^{K}\chi^{k,g}_{d}C^{k}(\vec r,t),\quad t\in [0, T], \vec r \in\Omega,\vspace{.1cm}\\
\phi^{g}(\vec r,t) =& 0 \quad\text{ on the boundary of the reactor core} : t\in [0, T], \vec r \in \partial \Omega ,\vspace{.1cm}\\
\phi^{g}(\vec r,0)=&\phi^{g}_0(\vec r) \quad\text{ the initial condition} : \vec r \in\Omega.
\end{array}\right.
\end{equation}
The delayed neutron concentrations satisfy:
 \begin{equation}\label{eqPREC}
 \dfrac{\partial C^{k}}{\partial t} (\vec r,t)= -\lambda_{k}C^{k}(\vec r,t)+\displaystyle\sum_{g'=1}^{\hat g}\beta^{k,g'}\nu^{g'}\Sigma_{f}^{g'}\phi^{g'}(\vec r,t),\quad  t\in [0, T],  \vec r\in\Omega.
 \end{equation}
In equations~\eqref{eqBLZM} and \eqref{eqPREC}, the neutron velocity $V^{g} = \sqrt{2E^g/m}$,  the diffusion coefficient is denoted by $D^{g}$, $\Sigma_{t}^{g}$ , $\nu\Sigma_{f}^{g}$ are the total and production cross-section, $\Sigma_{s}^{g'\rightharpoonup g}$ stands for the transfer cross-section from energy group $g'$ to $g$. The fission spectra of prompt and delayed neutrons, respectively $\chi^{g}_p$ , $\chi^{k,g}_{d}$. The concentration of precursor group is denoted by $C^{k}$, their delay fraction and decay constant are denoted by $\beta^{k,g}$ and $\lambda_{k}$ respectively. The total delay fraction is denoted by $\beta^{g}$, where $\beta^{g}=\displaystyle\sum_{k=1}^K\beta^{k,g}$. For further physical explanation and analysis of the model we refer to~\citep{BerirhanPhD} and reference therein.

 We denote, hereafter, by {\it reactivity} the energy of the reactor core equal to the sum over $g$ of the squared  $L^2(\Omega)$-norms of the fluxes  $\phi^{g}$, solutions of the neutron model  \eqref{eqBLZM}.	The reactivity of the reactor core remains a function of  time. Its evolution is essentially caused by the chain reaction of the neutrons' fission.
	In the reactor core this fission chain reaction produces exponentially in time new neutrons, responsible of the variation of the reactivity of the reactor. In order to control this effect, some rods that have the property of absorbing neutrons are sequentially  immersed and withdrawn  inside the reactor core. This action ensures the safety of the reaction during the production of the electricity.
	
%	The reactor core may have three states of energy: \textit {subcritical, critical and supercritical}. Those states correspond to the density of neutrons diffusing in the reactor. The reactor is said to be {\it critical} if the neutron outcome, in the sense of energy, is null. That means, the production and absorption of neutron are equal during time. The neutron outcome in the reactor at equilibrium obeys: 
%\begin{equation*}
%\hbox{production}_{fission} = \hbox{absorption}+\hbox{leaks}.
%%+\hbox{source}_\textit{external}
%\end{equation*}
%Neutrons injected into the system initiate a chain reaction, absorption could be important, due to absorber rods, and the leaks represents the neutrons leaving the system.\\
% The multiplicative factor in a medium without leaks, denoted by $k_{\infty} $ is defined by:
% \begin{equation*}
% k_{\infty}:=\dfrac{\hbox{neutron density at time $t_i$}}{\hbox{neutron density at time $t_{i-1}$}}.
%\end{equation*}
%In a bounded domain, the effective multiplicity factor $k_\text{eff}$ is defined by:
%\begin{equation*}
%k_\text{eff}=\dfrac{k_{\infty}}{1+\hbox{leaks}}.
%\end{equation*}
%This coefficient should be very close to unity in order to guarantee equilibrium of the energy in the reactor. Thus, if $k_ \text{eff}<1$ the state of the reactor  is said {\it subcritical}, else if $k_ \text{eff}> 1$ the state is {\it supercritical} else (i.e. $k_ \text{eff} =1$) the state is critical. 
%
The simulation of the neutron model generally starts from an equilibrium average flux distribution, which is characterized by a steady solution of~\eqref{eqBLZM}  where the unique source of neutron considered is the fission and we shall disregard the contribution of the delayed neutron i.e. $\beta^{g}$ and $\chi^{k,g}_d$ are assumed vanishing. Such an initial condition for the simulation of the neutron model is a solution to the  following eigenvalue problem, which looks for  a scalar $k_ \text{eff}$ known as the {\it effective reactivity}. This factor is linked to the largest generalized eigenvalue associated to the two operators: {\it production} operator, namely, ${\bf F}$ and {\it absorption} operator, namely, ${\bf P}$. The production operator characterizes the production of the neutrons by fission and is extracted from the fission source term $\chi^{g}_p\sum_{g'=1}^{\hat g}\nu^{g'}\Sigma_{f}^{g'}\phi^{g'}(\vec r,t)$, whereas absorption operator characterizes escapes or leaks in the system and corresponds to the diffusion and scattering term : $-\text{div}\big(D^{g}\nabla\phi^{g}(\vec{r},t)\big)+\Sigma_{t}^{g}\phi^{g}(\vec{r},t)-\displaystyle\sum_{g'=1}^{\hat g}\Sigma_{s}^{g'\rightharpoonup g}\phi^{g'}(\vec{r},t)$. The eigenvalues problem reads : find $\Lambda\in {\mathbb R}$ and $\Phi_\Lambda$ such that
$${\bf P}\Phi_\Lambda=\Lambda {\bf F}\Phi_\Lambda.$$
The solution $\Phi_{\Lambda_{max}}$ associated to  the largest eigenvalue  $\Lambda_{max}$ to this problem 
 is chosen as the initial condition to problem  \eqref {eqBLZM}, and we set $k_\text{eff} =
\dfrac{1}{\Lambda_{max}}$. The associated $C_{\Lambda_{max}}$ which corresponds to $\Phi_{\Lambda_{max}}$ through the steady state of \eqref{eqPREC} is chosen as the initial condition for the precursor concentrations.

Problem \eqref{eqBLZM} and \eqref{eqPREC}  can be written in a condensed form as follows:
\begin{equation}\label{cauchy}
\left\{\begin{array}{lll}
\dfrac{\partial y(t)}{\partial t} = {\bf A}(t)y(t),&\quad\text{on } [0,T]\times\Omega\\
y(t=0) =y_0,&\quad\text{on } \{0\}\times\Omega,
\end{array}\right.
\end{equation}
where $y$ stands for the solution $y:=(\phi^{1},\dots, \phi^{\hat g},C)^T$. In the case where $\hat g=2$ the reaction diffusion operator ${\bf A}(t)$ thus reads
$$\scriptsize 
{\bf A}(t) =\left(\begin{array}{c|c|c}
  V^{1}\text{div}(D^{1} {\nabla}(\bullet))-\Sigma_{t}^{1}(\bullet) + V^{1}\chi^{1}_p (1-\beta^{1})\nu^1\Sigma_{f}^{1}(\bullet) +V^{1}\Sigma_{s}^{(1\rightharpoonup 1)}(\bullet) & V^{1}\chi^{1}_p (1-\beta^{2})\nu^2\Sigma_{f}^{2}(\bullet) +V^{1}\Sigma_{s}^{(2\rightharpoonup 1)}(\bullet)& V^{1}\sum_{k=1}^{K}\chi^{k,1}_{d}(\bullet) 
\\\hline V^{2}\chi^{2}_p (1-\beta^{1})\nu\Sigma_{f}^{1}(\bullet) +V^{2}\Sigma_{s}^{(1\rightharpoonup 2)}(\bullet) &
  V^{2}\text{div}(D^{2}{\nabla}(\bullet))-\Sigma_{t}^{2}(\bullet) + V^{2}\chi^{2}_p (1-\beta^{2})\nu^2\Sigma_{f}^{2}(\bullet) +V^{2}\Sigma_{s}^{(2\rightharpoonup 2)}(\bullet) 
   & V^{2}\sum_{k=1}^{K}\chi^{k,2}_{d}(\bullet) 
\\\hline \beta^{k,1}\nu^1\Sigma_{f}^{1}(\bullet) & \beta^{k,2}\nu^2\Sigma_{f}^{2}(\bullet) & -\lambda_{k}(\bullet)\end{array}\right).
$$
The dependence in time of the operator ${\bf A}(t)$ is due to the possible change of cross sections values where the reaction occurs.
The existence of a solution to this problem is proven in \citep[chap  XXI, section 2]{MR792484}. It  can be written thanks to a   flow map as
follows 
\begin{equation}\label{2K10}
\forall t\ge 0, \forall\tau>0,\quad y(t+\tau)=\mathcal{E}_{\tau}^{t}(y(t)),
\end{equation}
where the uniqueness of the solution provides the semi-group property of the propagator $\mathcal E$, i.e. $$\mathcal{E}_{\tau'}^{t+\tau}\circ \mathcal{E}_{\tau}^{t} = \mathcal{E}_{\tau+\tau'}^{t}.$$

\section{The parareal in time algorithm}
%------------------------------------------------------------------------------------------
%									The parareal method
%------------------------------------------------------------------------------------------
	The parareal in time algorithm~\citep{MR1842465}  is a ``divide and conquer'' method that enables parallelization across the time direction. Following the same strategy as  in domain decomposition methods, the parareal in time algorithm is based on breaking up the time  interval of simulation into sub-intervals and solve  over each sub-interval independently using different processors by updating iteratively the initial condition over each sub-interval. The time evolution problem is thus broken up into a series of independent evolution problems on smaller time intervals. 
The parareal in time algorithm can be presented  as a predictor corrector process~\citep{ PhysRevE.66.057701, MR1962689}, and also as a multi-shooting algorithm also as a kind of Newton method with a time coarse grid defining the Jacobian matrix~\citep{ MR2334115}. Many improvements on the method, in particular for efficient iterative solution procedure on parallel architectures have been proposed, see e.g. ~\citep{MR2765386,Aubanel:2011:STP:1967754.1967824}.

Starting from the general formulation of  \eqref{cauchy}, we assume given a uniform partition of the time interval into $N$ subintervals, such that $0=t_0< t_1<..<t_n<t_{n+1}<..<t_N=T$, and denote $\Delta t=t_{n+1}-t_n$, so $t_n=n\Delta t$. Based on the semigroup property stated in \eqref{2K10}
we have  
$$
y(t_{n+1}) = \mathcal E_{\Delta t}^{t_n}(y(t_n)) = \mathcal E_{(n+1)\Delta t}^{t_0}(y_0)\quad\forall\, 0\leq n\leq N-1.
$$ 
In practice, we have to provide a fine enough numerical approximation of the propagator $\mathcal E$ denoted by $\mathcal F$ and named by fine propagator in what follows. It is based on an appropriate classical discretization scheme. In order to understand the mechanism of the parallelization across the time, let us denote by $ Y_n$ such a fine approximation to
the solution of the Cauchy problem \eqref{cauchy} at time $t_n$ i.e. $Y_n\simeq y(t_n)$. The sequence of solution $( Y_n)_{n=1}^{n=N-1}$ is thus solution of:
$$
 Y_{n+1}= \mathcal{F}_{\Delta t}^{t_n}( Y_n), \quad\forall\, 0\leq n\leq N-1\quad\text{with}\quad Y_{0}=y_{0} ,
$$
putting in evidence the sequential nature of the time propagation and, correlatively the fact that we are constrained to know first the actual (or approximated) solution e.g. at time $t_n$ in order to compute the solution at time $t_{n+1}$ (or its approximation). The parareal in time algorithm involves another propagator, denoted as $\mathcal G$ which is a coarse version of the fine propagator $\mathcal F$. The propagator $\mathcal G$ is assumed to be faster and cheaper, in order to be able to accomplish rapidly the sequential propagation of the solution from time $t_0$ to $t_N$, of course it cannot be so accurate as $\mathcal F$.  This allows to define a sequence of approximate solutions $(Y_n^k)^{k>0}$ that converges to the right solution $Y_n$ when $k$ goes to infinity. Starting from (at $k=0$)
\begin{equation}\label{parareal_init}
 Y_{n+1}^{0}=\mathcal G_{\Delta t}^{t_n}( Y_{n}^{0}), \quad\forall\, 0\leq n\leq N-1, \quad Y_0^0 = y_{0},
\end{equation}
the numerical scheme of the parareal method is  given by: (from $k$ to $k+1$) knowing $( Y_{n}^{k})_n$, compute $( Y_{n}^{k+1})_n$ by the following
\begin{equation}\label{parareal}
 Y_{n+1}^{k+1}=\mathcal G_{\Delta t}^{t_n}( Y_{n}^{k+1})+\mathcal F_{\Delta t}^{t_n}( Y_{n}^{k})-\mathcal G_{\Delta t}^{t_n}( Y_{n}^{k}),\quad\text{with } Y_{0}^{k}=y_{0}.
\end{equation}

The first contribution of the right hand side in \eqref{parareal} is the prediction for all propagated solution at time $t_{n+1}$, where the rest of the right hand side corresponds to the correction scheme, using the accurate solution with the accurate propagator $\mathcal F$ subtracting the inaccurate solution predicted before using the coarse propagator $\mathcal G$. 
 
\section{Numerical method}
%------------------------------------------------------------------------------------------
%									Numerical method
%------------------------------------------------------------------------------------------

	Thanks to the property of symmetry in the reactor core the computational domain is reduced to one of its quarter. Consequently, we consider Neumann boundary condition at these artificial interfaces, that is $\dfrac{\partial\phi^{g}}{\partial \vec n}(\vec r,t) = 0$, where $\vec n$ is the outward normal of the domain. 

\subsection{Setting-up the initial condition}\label{initialconditionconstuction}
%------------------------------------------------------------------------------------------
%									Initial condition 
%------------------------------------------------------------------------------------------

As was stated previously, the critical state of the reactor core is derived from the steady solution of~\eqref{eqBLZM} without the contribution of the delayed neutrons :
\begin{equation}\label{pvp}
-\text{div}(D^{g}\vec{\nabla}\phi^{g})(\vec{r},t)+\Sigma_{t}^{g}\phi^{g}(\vec{r},t)-\displaystyle\sum_{g'=1}^{\hat g}\Sigma_{s}^{g'\rightharpoonup g}\phi^{g'}(\vec{r},t)=
\dfrac{1}{k_\text{eff}}\chi^{g}\sum_{g'=1}^{\hat g}\nu^{g'}\Sigma_{f}^{g'}\phi^{g'}(\vec{r},t).
\end{equation}
Practically, the maximum eigenpair  (from which the factor $k_\text{eff}$ is computed) is solved by the use of the power algorithm. 
\subsection{Fine propagation}
%------------------------------------------------------------------------------------------
%									Fine propagator numerical scheme 
%------------------------------------------------------------------------------------------

%------------------------------------------------------------------------------------------fine  propagator
Numerical discretization of the neutron equation is briefly presented in this subsection. Equation \eqref{eqBLZM} is composed by several time dependent partial differential equations of order two in space and order one in time for the flux distribution and $K$ ordinary differential equations for the concentration of precursors. We first give the space discretization of those equations and then  give the numerical scheme to approximate the time dependency.

 The numerical scheme, to approximate space variable, is based on the classical Galerkin of finite element type, where the domain $\Omega$ is meshed with tetrahedral elements. We use P1-Lagrange finite element for the average flux and P0-Lagrange finite element for both concentration of precursors and the physical parameters.
In the sequel, the space-time approximation of the global unknown $y(t_i)$,  is denoted by $y_i$.
We now assume the following partition of the time interval $[t_0,t_{N}]=\cup_{n=0}^{N-1}[t_n,t_{n+1}]$ where $[t_n,t_{n+1}]$ is also divided into $I$ subintervals of equal size denoted by $\tau$. Consequently we have $[t_0,t_{N}]=\cup_{n=0}^{N-1}\cup_{i=0}^{I-1}[t_{n,i},t_{n,i+1}]$. With the obvious notations, $t_{n,0}=t_n$ and $t_{n,I}=t_{n+1}$. In the sequel the double subscript is dropped and the use of the subscript $\underline i = (n,i)$ stands, from now on, for the index of any time $t_{\underline i}$, whereas the use of the subscript $n$ stands only for time $t_n=n\triangle t$. 

	 The evolution in time of the solution $y_{\underline i}$ is approximated using the $\theta-$scheme, which assumes at time $t_{{\underline i}+1}$ the solution $y_{\underline i}$ is known. The solution at time $t_{{\underline i}+1}$ is thus computed by solving $y_{{\underline i}+1}-y_{\underline i} = \tau\theta{\bf A}_{{\underline i}+1} y_{{\underline i}+1} + \tau(1-\theta){\bf A}_{\underline i} y_{\underline i}$, where $\theta\in(0,1)$, and ${\bf A}_{\underline i}$ represents the approximation matrix of the operator ${\bf A}(t_{\underline i})$ at time $t_{\underline i}$. Finite element matrix representation can be found in e.g.~\citep[chap.3]{PhDriahi}. The resolution in the forward time, with $\theta\neq 0$, requires matrix inversion at each time step, such that
\begin{equation}\label{C-N}
y_{{\underline i}+1}=\big(I-\tau\theta{\bf A}_{{\underline i}+1}\big)^{-1}\big(I+\tau(1-\theta){\bf A}_{\underline i}\big) y_{\underline i}.
\end{equation}
 The efficient choice of the parameter $\theta$ is discussed in the experimental part. 

	More technical discussion concerning the fine propagation, in addition to the classical discretization scheme presented above, is given in the sequel. 
	
		The numerical approximation of the transient Langenbuch-Maurer-Werner (LMW) benchmark application~\citep{LMWref} requires a particular attention of the control rods movement, where the cross-sections have to be interpolated in order to avoid oscillations and instabilities of the numerical solver. Similar cross-sections interpolation technique is applied as in~\citep{IKEDAHIDEAKI} for the neutron nodal expansion method. %with the use of the improved quasi-static for spatial neutron kinetics.  

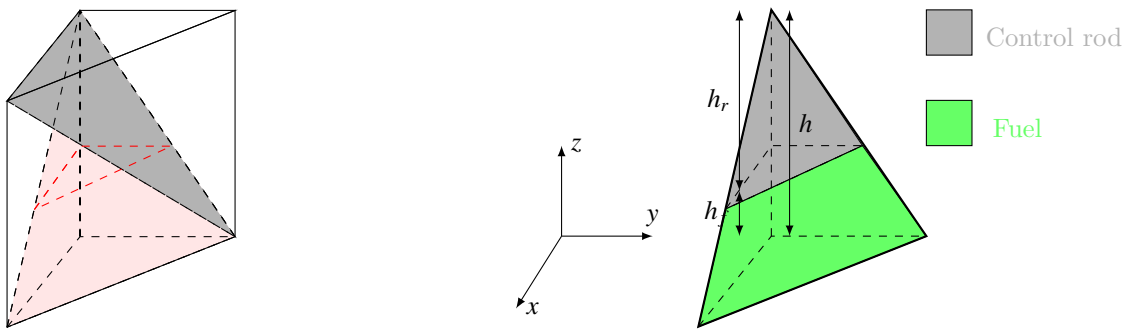
\begin{figure}[!h]\centering
\begin{minipage}[t]{0.4\linewidth}
  \begin{tikzpicture}[scale=1.2,>=latex]

   \begin{scope}
\draw[dashed,fill=red!10] (-.5,0) --(.3,3.5) -- (2,1) -- cycle ;
\draw[dashed,fill=gray!60]  (-.5,2.5) --(.3,3.5) -- (2,1) -- cycle ;
     \draw (-.5,0) -- (-.5,2.5)  -- (2,3.5) -- (2,1)-- cycle ;
     \draw (-.5,2.5) --(.3,3.5) -- (2,3.5) -- cycle ;
     \draw[dashed] (-.5,0) --(.3,1) -- (2,1) -- cycle ;
     \draw[dashed] (.3,1) -- (.3,3.5) --cycle ;
          \draw[dashed] (-.5,0) --(.3,3.5);
      \draw[dashed] (2,1) --(.3,3.5);
      \draw[dashed] (-.5,2.5) -- (2,1);
    \end{scope}
    \begin{scope}
          \draw[dashed,color=red] (-0.2,1.3) -- (0.3,2) -- (1.3,2) --  cycle ;
          \draw[dashed,color=red] (-0.2,1.3) -- (0.3,2)  ;
    \end{scope}
    \end{tikzpicture}\end{minipage}
\begin{minipage}[t]{0.4\linewidth}
\pgfdeclareverticalshading{myshadingG}{100bp}
{color(0bp)=(red); color(25bp)=(green!60); color(75bp)=(blue!60); color(100bp)=(black)}
%\centering
{
  \begin{tikzpicture}[scale=1.2,>=latex]
\begin{scope}
\draw[fill=gray!60] (-0.2,1.3) -- (.3,3.5) -- (1.3,2) --  cycle ;
\draw[fill=green!60] (-.5,0)  -- (2,1)  -- (1.3,2) -- (-0.2,1.3) -- cycle;
\end{scope}
   \begin{scope}[thick]
   \draw (-.5,0)  -- (2,1) -- (0.3,3.5)-- cycle;
   \end{scope}
   \begin{scope}[dashed]
\draw (-0.2,1.3) -- (0.3,2) -- (1.3,2) --  cycle ;
   \draw (.3,1) -- (.3,3.5);
   \draw (-.5,0) --(.3,1) -- (2,1) -- cycle ;
   \end{scope}
\begin{scope}
\pgfpathmoveto{\pgfpointorigin}
\pgfpathlineto{\pgfpoint{0cm}{0cm}}
\pgfpathlineto{\pgfpoint{.3cm}{1cm}}
\pgfpathlineto{\pgfpoint{2cm}{1cm}}
\end{scope}
\begin{scope}[->]
\draw (-2,1) -- (-1,1) node[above] {$y$};
\draw (-2,1) -- (-2.5,0.2) node[right] {$x$};
\draw (-2,1) -- (-2,2) node[right] {$z$};
\draw[<->] (.5,1) -- (.5,3.5) node[midway,right]{$h$};
\draw[<->] (-.05,1.5) -- (-.05,3.5) node[midway,left]{$h_{r}$};
\draw[<->] (-.05,1) -- (-.05,1.5) node[midway,left]{$h_{f}$};
\end{scope}
\begin{scope}
\draw[fill=gray!60]  (2,3) -- (2,3.5) -- (2.5,3.5) -- (2.5,3) --  cycle;
\node at (3.4,3.2) [color=gray!60] {Control rod};
\draw[fill=green!60] (2,2) -- (2,2.5) -- (2.5,2.5) -- (2.5,2) --  cycle ;
\node at (3,2.2) [color=green!60] {Fuel};
\end{scope}
  \end{tikzpicture}
}
\end{minipage}
\caption{ P0-interpolation of the cross-sections over one simplex (tetrahedron). The triangulation of the domain $\Omega\subset\mathbb{R}^3$ is structured in the sense that each tetrahedron has one vertical edge, suppose that $h$ is the hight of one tetrahedron we denote by $h_r$ and $h_f$ the length of the penetration of control rods in the tetrahedron and the rest of the length respectively. The distance $h_r$ and $h_f$ are therefore perfectly determined using the velocity of control rods.}\label{projectP0}
\end{figure}

	Cross sections are approximated with piecewise constant functions. In order to take into account their unsteadiness resulting from the immersion of control rods within the reactor, one could refine the mesh in order to follow the movement of the rods. This would be quite expensive, another possibility would be to tune the time step with the velocity of the movement so that the rod ends exactly match the mesh. Of course this adds strong constraint on the design of the discretization. In order to cope with this, we assume that   the triangulation of the domain is extracted from a first decomposition of $\Omega$  into several horizontal layers of prisms each of them being subsequently cut into three tetrahedrons as explained in Figure~\ref{projectP0}(left). This enables interpolating the cross sections, as sketched in Figure~\ref{projectP0}(right) where the immersion of the control rod into the tetrahedron is illustrated. The cross section (represented with color) are taken into account relatively with the volume occupied by control rod. In this case positive values, between $0$ and $1$, is attributed to the cross section related to the rods, whereas cross section related to fuel are deduced automatically i.e. interpolated using the complement value. It is worthy noting that this procedure is done before finite element matrix assembly. We update cross sections at each time step using interpolation, then we assemble the main matrix and solve the system of~\eqref{C-N}.        

	  %------------------------------------------------------------------------------------- begin tikz picture 
	  %------------------------------------------------------------------------------------- end tikz picture

%------------------------------------------------------------------------------------------coarse propagator

\subsection{Coarse propagation using reduction of the model}\label{subsec:Coarse}
	In this subsection we present and discuss a reduction of the previous model; this simplified system will be used in the parareal in time algorithm as a coarse solver for the simulation in order to further diminish the cost of the coarse sequential solver. The simplification we use deals with  the number of groups used both for flux and precursors. The reduced system has thus less number of unknown (i.e. just \eqref{eqBLZM} without coupling with precursor \eqref{eqPREC}). Hence, we reduce the numerical system and accelerate its resolution. We propose an additional simplification  by reducing the motion of control rods to a two specific positions. The resulting reduced motion is called {\it static}, which refers to a fixed control rods positioning, whereas, the continuous motion of rods is called {\it dynamic}.  This allows to minimize the complexity of updating the matrices according to the evolution of control rods position. 
	
Note however that some care has to be taken so that the reduced system remains coherent with the exact dynamics. Indeed the introduction of the effectivity factor $k_\text{eff}$ allows to tune the equations and further define \textit {subcritical, critical and supercritical} states. Of course the movement of the rods changes the state of the core. By modifying the coarse model as we did changes rather completely this equilibrium, first by the fact that  the  delayed neutrons have a very important role in the stability and the damping of the constantly increasing energy produced by fission. The contribution of the concentrations of precursors is fundamental to access accurate information about flux density at time $t$ during the reaction. Hence the absence of the delayed neutrons in the reaction behavior inside the reactor core reduces the damping and energy production is accelerated . The reaction of the reduced model is more rapid in time, so we are able to reach, approximatively in time and quantity, the peak of the energy, one has to manipulate the velocity of neutrons in this case. Further, by freezing the position of the rods modifies also the reactivity of the model. In order to get the coarse solver closer to the full one, we tune the  reactivity factor $k_\text{eff}$ to two values from their critical value. Indeed the first period of the benchmark corresponding to a supercritical nature is represented by a reactivity factor slightly larger than the $1/\Lambda_{max}$ whereas the second stage of the benchmark, associated with the rod going down, is represented by a reactivity factor slightly smaller than the $1/\Lambda_{max}$.

The motion of control rods is therefore reduced, and the resulting reduced model produces the same state of the reactor before and after the peak. This reduction enables us to make computation cheaper while keeping accurate results as we show it in the experimental part later on (see section~\ref{exp-part}).

\subsection{The algorithm}
We have actually the ingredients to start up our algorithm, which we present with a frame on {\bf Algorithm}~\ref{algopararealkinetic}.The time parallel algorithm is implemented with a master-slave configuration for which we have two types of communications: a {\it distributive} and a {\it collective} communication. In the {\it distribution} communication, the main processor sends information towards all its processors agents. On the other hand in the {\it collection} communication; the Master himself receives and collects information since his agents. In both cases, it is about the same quantity of informations which passes in the two directions. The second type of communication is devoted to the correction of the coarse error, which requires fine information to be communicated by agent processors toward master.
	 We use some keywords from the parallel programing language to describe communication procedures. For instance, {\bf Recv}({\it data, sender}) and {\bf Send}({\it data,receiver}) mean that the processor who execute those commands receives the {\it data} from the {\it sender} and send {\it data} to the {\it receiver}. Whereas the {\bf Broadcast}({\it sender,data}) means that the processor {\it sender} send {\it data} to all other processors. 
\begin{algorithm}[phtb!]
\footnotesize
\label{algopararealkinetic}
%\SetVline\SetLine%[top=2in, bottom=4in, left 8in, right 0.2in]
\SetAlgoLined
\caption{Parareal kinetic of neutrons}
\SetKwFunction{KwMs}{\footnotesize {\bf Send}}
\SetKwFunction{KwMr}{\footnotesize {\bf Recv}}
\SetKwFunction{KwMb}{{\footnotesize \bf Broadcast}}
\SetKwFunction{KwCl}{\footnotesize {\bf  Call:}}
\SetKwFunction{KwRt}{\footnotesize {\bf Routine}}
\SetKwFunction{KwEd}{\footnotesize {\bf  end}}
\KwIn{ $N:=\# slave\, proc$, $\tau_{F}$,$\tau_{G}$}
\KwIn{${Y}_{0}^{0}=[(\Phi_{\Lambda_{max}})^T,({\bf c}_{\Lambda_{max}})^T ]^T$ as initial conditions, $\epsilon^\infty$ a tolerance of the algorithm \;}
\KwIn{ Solver $\mathcal A$, data vector $y$\;}
\KwRt{$\mathcal{ A},{y}$}
{
	 \begin{enumerate}[1)]
	\item Positioning the absorber rods with respect to their dynamic chronology\; 
	\item Assembling matrices related to Eqs.~\eqref{eqBLZM}-\eqref{eqPREC}\;
	\item Solving iteratively in time the system of~\eqref{C-N} (the result is denoted by $\mathcal A{y}$)\;
 	\end{enumerate}
}
\KwEd\KwRt\;
$k\longleftarrow 0$\;
{\Repeat
{$\max_{n}\epsilon_{n}^k\leq {\epsilon^\infty}$}
{	\eIf{master processor}
	{	\ForEach{$n\in\{0,.., N-1\}$}{
	\begin{enumerate}[1)]
	\item \KwCl \KwRt{${\mathcal G}_{\Delta t}^{t_n},{Y}_{n}^{k}$ } (i.e. coarse-serial propagation)\label{item1}\;
	\item$ $\\
	\eIf{$k=0$}
	{
	\lRepeat{$n=N-1$}{{\bf return to}~\ref{item1}{ \bf with } ${Y}_{n+1}^k$}
	}{
	Construct $({Y}_{n}^k)_{n\geq1}^{}$ with respect to the algorithm~\eqref{parareal}\label{item3}\;
	}
	\item \KwMs (${Y}_{n}^{k}, processor ( n )$) \label{item4}\;
 	\end{enumerate}
	}}
	{
	\ForAll{$slave\, processor (n) /n\in\{0,\hdots,N-1\}$}
	{
	\KwMr (${Y}_{n}^{k},master\, processor$)\;
	\KwCl \KwRt{ ${\mathcal F}_{\Delta t}^{t_n},{Y}_{n}^k$ } (i.e. fine-parallel propagation)\;
	\KwMs (${\mathcal F}_{\Delta t}^{t_n}{Y}_{n}^k, processor ( n ) $) \;
	 }
	 }
	\If{master processor}
	{
 	\ForEach{$n\in\{0,.., N-1\}$}{
	\KwMr (${\mathcal F}_{\Delta t}^{t_n}{Y}_{n}^k, processor( n )$) \;
	 Evaluate $\epsilon_{n+1}^k=\|{\mathcal F}_{\Delta t}^{t_n}{Y}_{n}^{k}-{Y}_{n+1}^{k}\|_{2}^2\slash\|{Y}_{n+1}^{k}\|_{2}^2$\;
	}
	}
 	$ k\leftarrow k+1$\;
	\KwMb ($master\, processor,\epsilon_{n}^k$)\;
}
}
\end{algorithm}

\section{Experimental part}\label{exp-part}

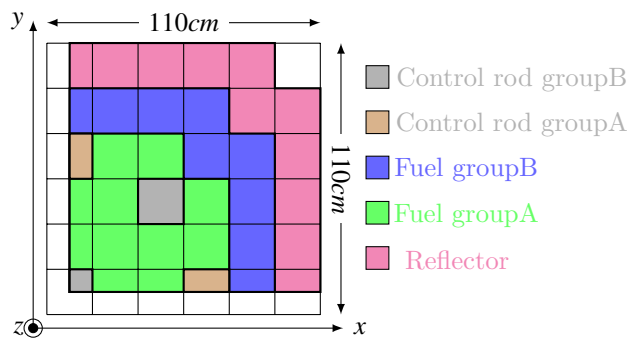
\begin{figure}[htpb]\centering
%\begin{minipage}{0.49\linewidth}
%\includegraphics[width=8cm,height=6cm]{./figure/Benchmark_LWR_Geometries}  
%\end{minipage}
%\begin{minipage}{0.49\linewidth}
%  \includegraphics[width=8cm,height=6cm]{./figure/mesh_3d.png}
%\end{minipage}
\begin{tikzpicture}[scale=6,>=latex]
\begin{scope}[thick]
\draw[fill=magenta!60]     (0.05,0.05) -- (.6,0.05) -- (.6,.5) -- (.5,.5) -- (.5,.6) -- (0.05,.6) -- cycle; 
\draw[fill=blue!60] (0.05,0.05) -- (.5,0.05) -- (.5,.4) -- (.4,.4) -- (.4,.5) -- (0.05,.5) -- cycle; 
\draw[fill=green!60] (0.05,0.05) -- (.4,0.05) -- (.4,.3) -- (.3,.3) -- (.3,.4) -- (0.05,.4) -- cycle; 
\draw[step=.1,black,thin,xshift=0.cm,yshift=0.cm]  (0,0) grid (.6,.6);
\draw[fill=gray!60] (.05,.05) rectangle (.1,.1);                         % rod 2
\draw[fill=gray!60] (.2,.2) -- (.3,.2) -- (.3,.3) -- (.2,.3) -- cycle; % rod 2
\draw[fill=brown!60] (.3,.05) -- (.3,.1) -- (.4,.1) -- (.4,.05) -- cycle; % rod 1
\draw[fill=brown!60] (.05,.3) -- (.1,.3) -- (.1,.4) -- (.05,.4) -- cycle; % rod 1
\end{scope}
\begin{scope}
\draw[fill=gray!60]      (.7,.5) -- (.7,.55) -- (.75,.55) -- (.75,.5) --  cycle; \node at (1.02,.52) [color=gray!60] {Control rod groupB};
\draw[fill=brown!60]   (.7,.4) -- (.7,.45) -- (.75,.45) -- (.75,.4) --  cycle; \node at (1.02,.42) [color=gray!60] {Control rod groupA};
\draw[fill=blue!60]          (.7,.3) -- (.7,.35) -- (.75,.35) -- (.75,.3) --  cycle; \node at (.92,.32) [color=blue!60] {Fuel groupB};
\draw[fill=green!60]   (.7,.2) -- (.7,.25) -- (.75,.25) -- (.75,.2) --  cycle; \node at (.92,.22) [color=green!60] {Fuel groupA};
\draw[fill=magenta!60]       (.7,.1) -- (.7,.15) -- (.75,.15) -- (.75,.1) --  cycle; \node at (.9,.12) [color=magenta!60] {Reflector};
\end{scope}
\begin{scope} % repere xyz
\draw[->] (-.03,-.03) -- (.65,-.03) node[right]{$x$};
\draw[->] (-.03,-.03) -- (-.03,.65) node[left]{$y$};
\draw (-.03,-.03) circle [radius=.02];
\filldraw (-.03,-.03) circle [radius=.01,color=black] node[left]{$z$};
\end{scope}
\begin{scope}
\draw[<->] (.645,0) -- (.645,.6) node[midway,fill=white,rotate=-90]{$110cm$};
\draw[<->] (0,.645) -- (.6,.645) node[midway,fill=white]{$110cm$};
\end{scope}
\end{tikzpicture}
%\end{minipage}
\caption{LMW transient problem: cross-section configuration in regard to rods positioning. Horizontal cross-sections .}%on the left and 3d cross-sections on the right.}
\label{fig:LMW}
\end{figure}

	In this section, we simulate the Langenbuch-Maurer-Werner (LMW) benchmark application~\citep{LMWref}, which is a kinetic benchmark in regard to the effect of delayed neutron fraction on the numerical transient application. The LMW simulates the kinetic of neutrons~\eqref{eqBLZM} involving rod movement dealing with a simplified large light water reactor. The LMW benchmark presented in Figure~\ref{fig:LMW}, as a quarter of the hole domain, initiates by withdrawing a bank of four partially inserted control rods inside the reactor core for certain time, then, another bank of five control rods is inserted. The global transient time is about $60$ sec when the velocity of all banks of control rods is about $3$ cm/sec. As explained before, the rods motions modeling has important discrepancy on the cross-sections that should be interpolated in order to avoid errors and undesired oscillations on the solution.
	We give in the appendix the cross-sections of the LMW transient problem and discuss experiments involving parallelization across the time for the numerical simulation.
	
\subsection{The parameters of the physic}

%------------------------------------------------------ Figure LWR Geometry

\begin{figure}
\begin{tikzpicture}

    \end{tikzpicture}
\end{figure}
	The motion of control rods in the LMW benchmark application reads: At time $t=0$ sec the first group of rods has an initial position at $z=100$ cm, while the second group has a higher position at $z=180$ cm. The velocity of the control rods motion is about $3$ cm/sec and is summarized as follows:
	\[\begin{array}{ll}
\hbox{Group 1 }: (t=0 \,|\, z=100\text{ cm})\nearrow(t=26,5  \,|\, z=180\text{ cm}),\\
\hbox{Group 2 }: (t=7,5 \,|\, z=180\text{ cm})\searrow(t=47,5 \,|\, z=60\text{ cm}).
	\end{array}\]

	 Cross-sections of the fuel (group-A and group-B), control rods  (group-A and group-B) and reflectors are given in Table~\ref{tab:tdn0} of the appendix. Scattering cross-section for the different medium are given in Table~\ref{tab:tdn1} of the appendix.
\subsection{Numerical tests and discussions}
	This section is dedicated to numerical tests of the LMW parallel in time transient problem. We outline our discussion into four steps: We first give simulation results related to the brute application of the parareal in time algorithm on the model described in section~\ref{sec:model}, next we present results related to the reduced model where we disregard delayed neutrons on the initial model. Then, we couple the two models. We finally discuss the speed-up achieved with the algorithm.

	 The numerical simulations were carried out on a parallel shared-memories-machine, which has $64$ processor with $2.0$ GHz, $256$ Go of shared memory and a communication network Numalink ($15$ GB/sec).  The parallelization of the procedure is carried out using  MPI library and the scientific computation software FreeFem++\citep{FFpp}.

	The analytical solution of~\eqref{eqBLZM} is not available, we produce thus a sufficiently approximated numerical solution using a refined time step. In what concerns space approximation, we use first order polynomial approximation with nodal P1-Lagrange finite element, where the domain is meshed with tetrahedron, since the main focus is about spacial error. %The refined numerical solution is, therefore, used as a reference for our parallel simulation. 
	
	As described above, the parareal in time algorithm present two types of solver that alternate, during the time process, between coarse-sequential resolution and fine-parallel resolution of~\eqref{eqBLZM}. The brut application of that algorithm need thus two type of numerical time-space discretization. In this paper we focus only on the coarse time discretization aspect and refer to~\citep{MR2235770} for an consideration of a coarse-space based solver with application to Navier–Stokes problem.   
	
\begin{figure}[!htbp]
  \begin{minipage}[b]{0.49\linewidth}
\includegraphics[width=8cm,height=5cm]{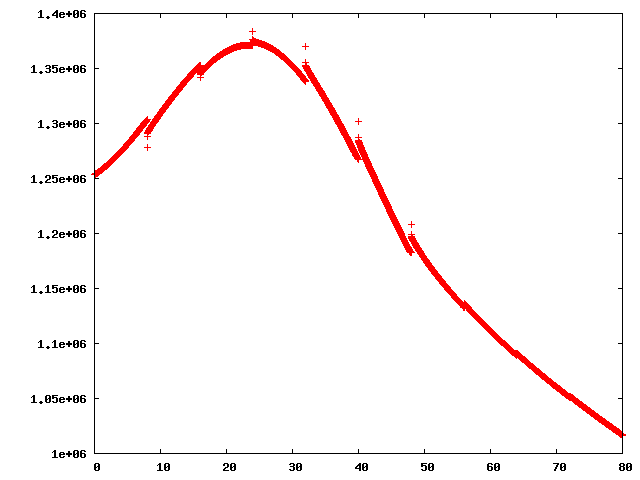}\\
\includegraphics[width=8cm,height=5cm]{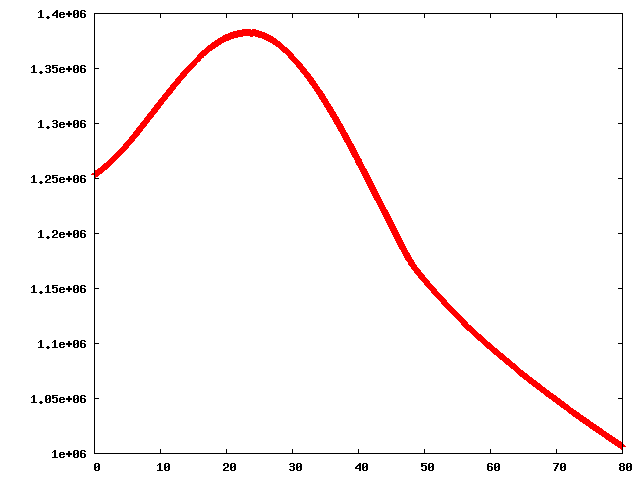}
\end{minipage}
  \begin{minipage}[b]{0.49\linewidth}
\includegraphics[width=8cm,height=5cm]{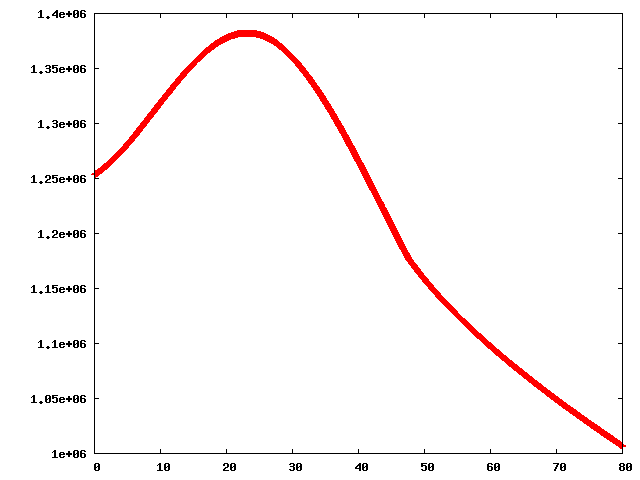}\\
\includegraphics[width=8cm,height=5cm]{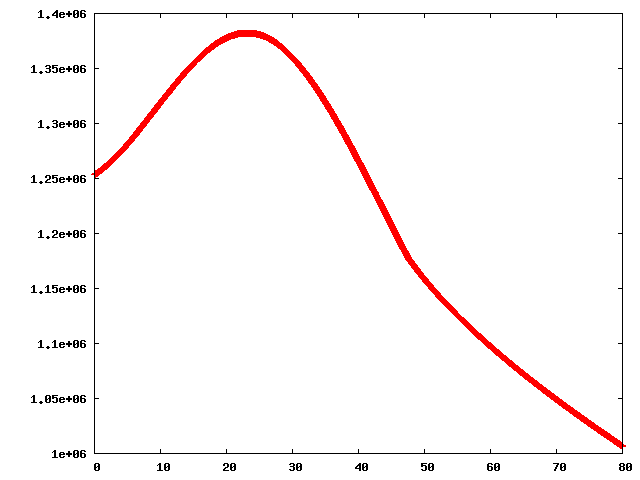}
\end{minipage}
\caption{First 4 iterations of the parareal in time algorithm for the LMW transient model: $\tau_{F}=10^{-1}$, $\tau_{G}=4$ et $N=10$.}\label{curvesc3}
\end{figure}

	We use $\theta$-scheme (see~\eqref{C-N}) in order to approximate the time variation of the solution, where $\theta=0$ leads to Euler Explicit scheme, $\theta=1$ leads to Euler Implicit scheme of order $1$. Numerical test showed that the case $\theta=1$ is the most stable scheme that provides an accurate solution without oscillation on its $L^2$ representation. We present in Figure~\ref{curvesc3} the $L^2$ trajectory of the flux distribution solution of the LMW produced with the parareal in time algorithm. 
The series of figures given in Figure~\ref{curvesc3} represent the first fourth iterations of the parareal algorithm. We remark that convergence of the trajectory occurs in few iterations using parallel simulation. These results show the evolution of the energy production with respect to the simulated time of the reaction (in seconds). The energy initiates with a value of $1.25e+6$ that represent the equilibrium of the reactor. We recall that bank of four control rods are already immersed in the reactor core at time $t=0$. Those rods are withdrawn simultaneously where after $7.5$ sec another group of control rods is immersed. The energy of the reactor (presented here as $L^2$ norm of the neutron flux distribution) is therefore achieving a peak before decreasing with the effect of the neutron absorption by the immersed rods.    
\begin{figure}[!hp]
\centering
  \begin{minipage}[b]{0.49\linewidth}
\includegraphics[width=8cm,height=5cm]{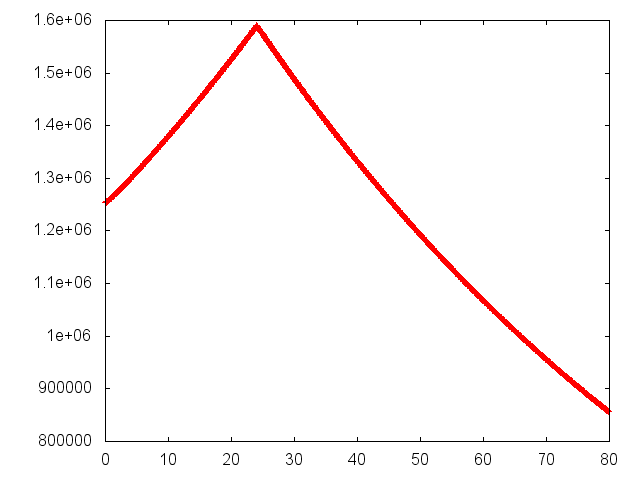}\\
\includegraphics[width=8cm,height=5cm]{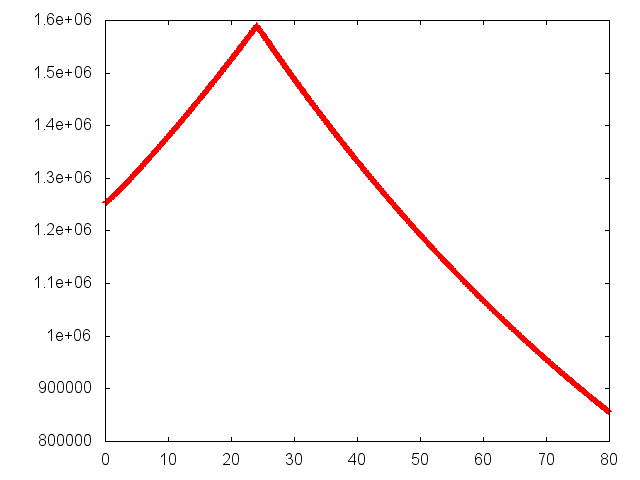}
\end{minipage}
  \begin{minipage}[b]{0.49\linewidth}
\includegraphics[width=8cm,height=5cm]{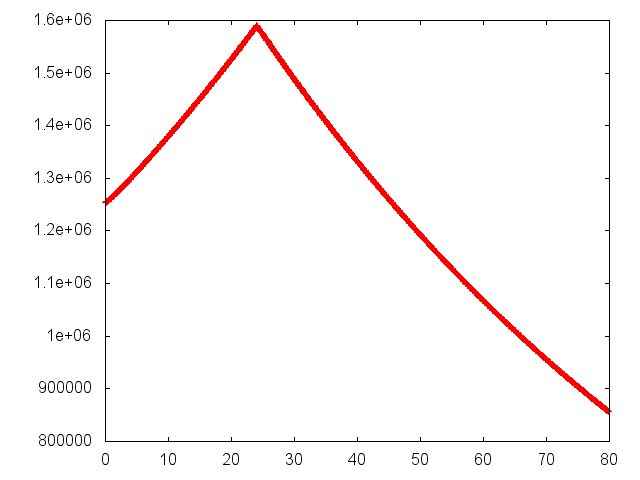}\\
\includegraphics[width=8cm,height=5cm]{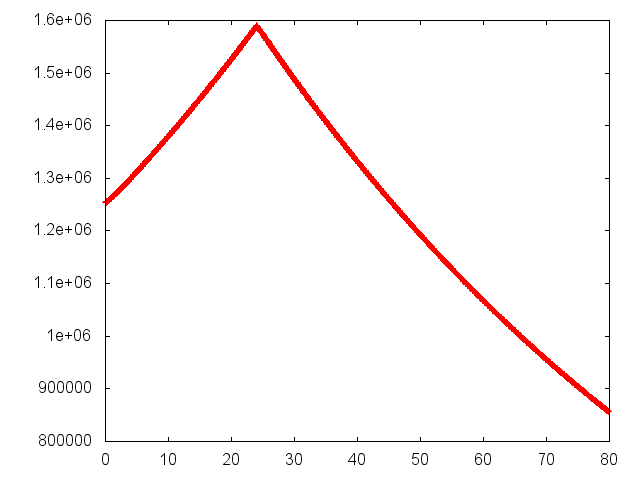}
\end{minipage}
\caption{First 4 iterations of the parareal in time algorithm of the reduced neutron model.$\tau_{F}=1$,  $\tau_{G}=4$ et $N=10$.}
\label{curvesc2}
\end{figure}
	
	The simulation of the reduced model is presented in Figure~\ref{curvesc2}. Here the parareal in time algorithm is used to produce parallel solution as the previous computation where the delayed neutrons don't contribute in the solution. We thus test and study sensitivity of the model as regards the reduction of the involved unknown and the brusque change of the position of control rods. As explained before in subsection~\ref{subsec:Coarse}; in the reduced model, we use change of reactivity instead of changing control rod positioning. This fact, avoid interpolating cross-sections at each position change. The global matrix is assembled only once where rods are fixed on their initial positions. Table~\ref{psg2} shows the corresponding change of the reactivity with respect to the control rods positions. 
\begin{table}[!hp]
\centering
	\begin{tabular}{ccc}\hline\hline
	 &$t\in[0,20] $& $t\in]20,80]$ \\\hline \hline
	 \hbox{Group 1} & z=160 cm&z=180 cm \\
	 \hbox{Group 2} & z=180 cm& z=60 cm\\\hline\hline
	 \hbox{reactivity} & 10008.e-5&9998.e-5 \\\hline
	 \end{tabular}
	 \caption{Change of the reactivity with respect to the control rods immersion.} \label{psg2}
\end{table}  

\begin{figure}[!hp]
\begin{minipage}[b]{0.48\linewidth}
\includegraphics[width=8cm,height=6cm]{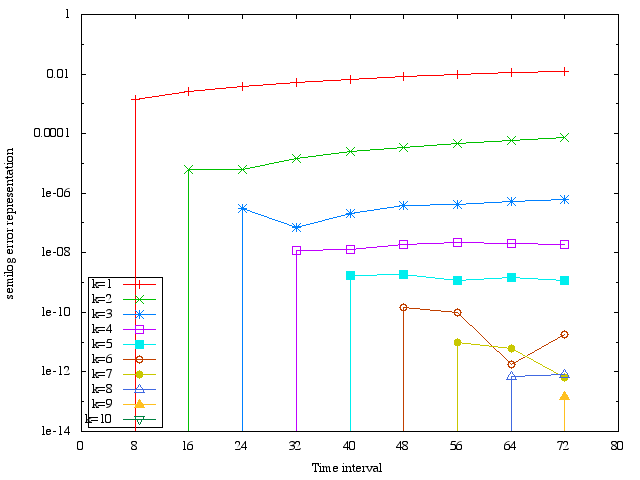}
\end{minipage}
\begin{minipage}[b]{0.48\linewidth}
\includegraphics[width=8cm,height=6cm]{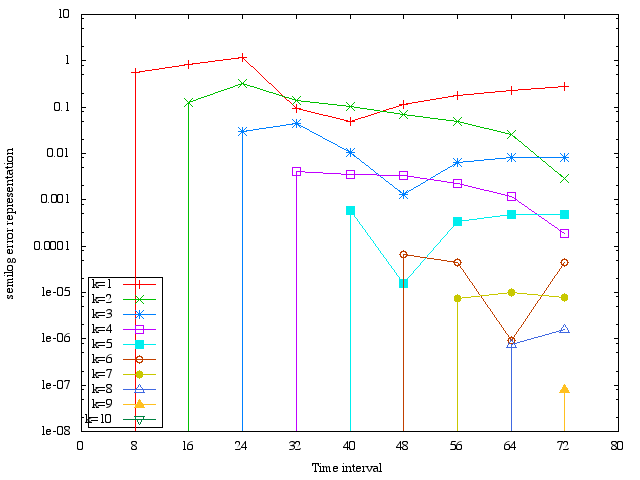}
\end{minipage}
\caption{Convergence of the algorithm (iteration 00-09) that couple the LWM transient model with the reduced model: (left)$\tau_{F}=10^{-1}$, $\tau_{G}=2$ and $N=10$. (right) $\tau_{F}=10^{-1}$, $\tau_{G}=4$ et $N=10$. Remark that errors lines appear at time $(k+1)T/N$ with vertical line, this occurs because error before that time is vanishing where parallel solution is exactly equal to the serial solution at time $t<(k+1)T/N$ }
\label{figcvg}
\end{figure}

	 Let us show numerical convergence results. We present in Figure~\ref{figcvg} the convergence of the algorithm with a threshold $\epsilon$ taken as the maximum among $\epsilon_{n}^k$, where :
$$
\epsilon^k:=\max_{n}\epsilon_{n}^k, \text{ where } \epsilon_{n}^k=\dfrac{\|{Y}_{n}^k - Y_{n}\|_2^2}{\|Y_{n}\|_2^2},
$$
where $Y_n$ is assumed to be the reference numerical solution, whereas $Y_n^k$ represents the numerical parallel solution produced with the parareal in time algorithm. 
		
	Taking into account the first order accuracy of the Euler-Implicit time-discretization scheme and the fact that the referred-to numerical solution (supposed to be very fine) reproduced with a time step  $10^{-2}$, we can thus fix the stopping criterion of the algorithm taking into account this information. 
	We consider the convergence of the parareal in time algorithm to an error $\epsilon^\infty$ of order $10^ {- 3}$. We complete the graphic results with Tables~\ref{tab:a1} and~\ref{tab:a2} presenting robustness of the methods with different choice of time-steps discretization. A thorough convergence results are as well given in the Tables.

\begin{figure}[!hp]
\centering
  \begin{minipage}[b]{0.49\linewidth}
\includegraphics[width=8cm,height=5cm]{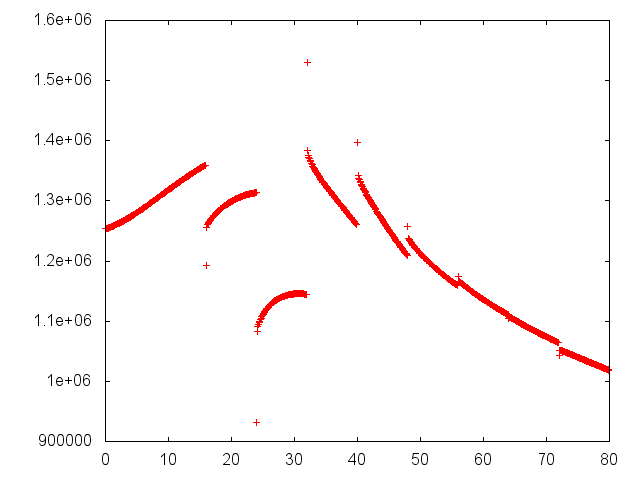}\\
\includegraphics[width=8cm,height=5cm]{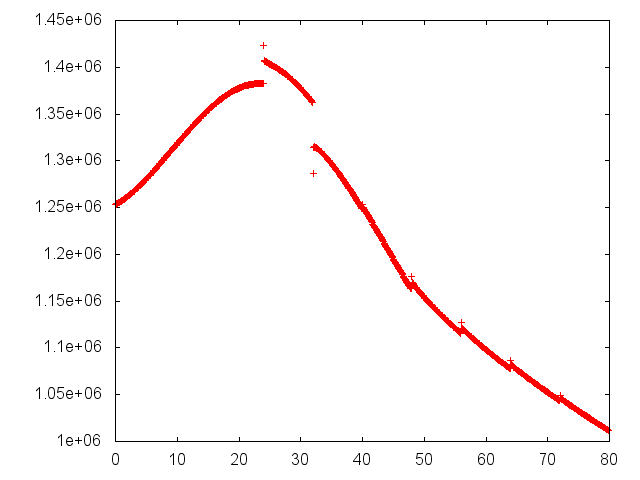}
\end{minipage}
  \begin{minipage}[b]{0.49\linewidth}
\includegraphics[width=8cm,height=5cm]{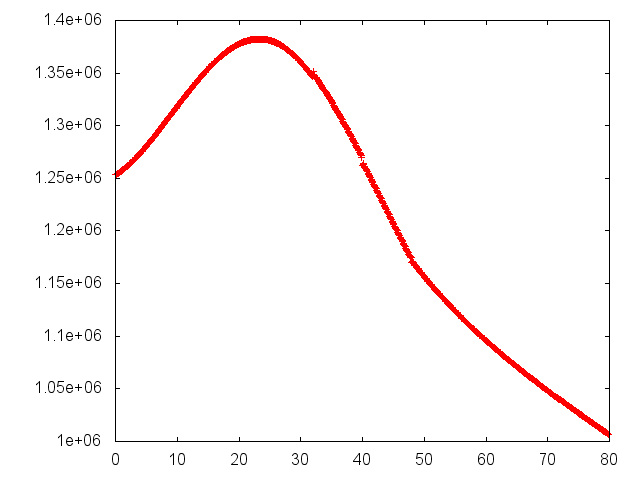}\\
\includegraphics[width=8cm,height=5cm]{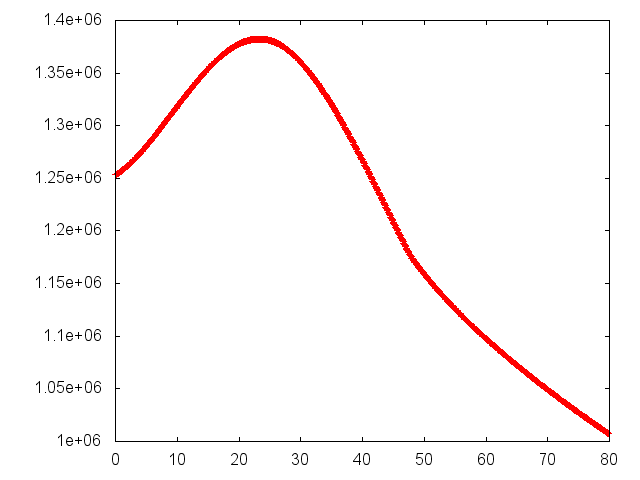}
\end{minipage}
\caption{First 4 iterations of the parareal in time algorithm for the LMW transient model with the reduced neutron model, coarse time step is also used for the reduced model: $\tau_{F}=10^{-1}$, $\tau_{G}=2$ and $N=10$.}\label{pow-sc2}
\end{figure}

Convergence results related to the combination of the the initial model and the reduced model, in the resolution by the parareal algorithm, are given in Figure~\ref{pow-sc2}. We recall that the coarse model, in addition of reducing the number of unknown, changes the reactivity of the system instead of change position of control rods. The employed procedure accelerates the computational time. Indeed, only one multiplication of the matrix (already in memory) by an adequate real coefficient is enough to reproduce the new matrix related to the operator ${\bf A}(\tau)$ at a desired time $\tau$. 
It is worthy noting, by the way, that the sequential simulation lasts {\bf 3 hours}.

%-------------------------------------------
\begin{table}[!htbp]
\vspace{.4cm}\centering
%   \begin{minipage}[t]{0.5\linewidth}\small\centering
\begin{tabular}{cccccc}\hline\hline
 $\tau_{F}$ & $\tau_{G}$&$\displaystyle\max_{n\geq0}\epsilon^1_{n}$&$\displaystyle\max_{n\geq0}\epsilon^2_{n}$&$\displaystyle\max_{n\geq0}\epsilon^3_{n}$&$\displaystyle\max_{n\geq0}\epsilon^4_{n}$\vspace{.05in}\\\hline\hline
0.01	&	2	&	1.42e-02	&	3.04e-05	&	1.73e-05	&	1.12e-06	\\
0.01	&	4	&	2.93e-02	&	1.16e-03	&	1.22e-04	&	1.46e-05	\\
0.01	&	8	&	6.09e-02	&	4.11e-03	&	7.85e-04	&	1.43e-04	\\
0.1	&	0.5	&	2.92e-03	&	1.31e-05	&	1.41e-07	&	2.08e-09	\\
0.1	&	1	&	6.53e-03	&	6.48e-05	&	1.67e-06	&	5.12e-08	\\
0.1	&	2	&	1.35e-02	&	2.81e-04	&	1.51e-05	&	9.22e-07	\\
0.1	&	4	&	2.88e-02	&	1.11e-03	&	1.13e-04	&	1.31e-05	\\
0.1	&	8	&	6.04e-02	&	4.02e-03	&	7.53e-04	&	1.35e-04\\
0.5	&	2	&	1.13e-02	&	1.71e-04	&	7.02e-06	&	3.35e-07	\\
0.5	&	4	&	2.68e-02	&	0.80e-04	&	7.80e-05	&	7.92e-06	\\
0.5	&	8	&	5.83e-02	&	3.55e-03	&	6.17e-04	&	1.02e-04	\\
1	&	2	&	7.58e-03	&	7.42e-05	&	1.94e-06	&	6.09e-08	\\
1	&	4	&	2.29e-02	&	6.30e-04	&	4.64e-05	&	3.91e-06	\\
1	&	8	&	5.43e-02	&	3.01e-03	&	4.77e-04	&	7.12e-05	\\
\hline\hline
\end{tabular}
\caption{Iteration $1,2,3$ and $4$ of the algorithm, complet model with dynamic rods scenario.}\label{tab:a1}
\end{table}
%%%%%%%%%%%%%%%%%%%%%%%%%%
\begin{table}[!htbp]\centering
\vspace{.2cm}
\begin{tabular}{cccccc}\hline\hline\small
 $\tau_{F}$ & $\tau_{G}$&$\displaystyle\max_{n\geq0}\epsilon^1_{n}$&$\displaystyle\max_{n\geq0}\epsilon^2_{n}$&$\displaystyle\max_{n\geq0}\epsilon^3_{n}$&$\displaystyle\max_{n\geq0}\epsilon^4_{n}$\vspace{.05in}\\\hline \hline
  0.1	&	2	&	1.39e-02 &	5.19e-05 &   	1.91e-07    &		5.71e-09	\\
0.1	&	4	&	2.78e-02 &    	1.88e-04 &	1.74e-06    &		1.52e-07	\\
0.1	&	8	&	5.34e-02 &	5.64e-04 &	1.63e-05    &		2.53e-06	\\
0.5	&	2	&	1.09e-02 &	3.17e-05 &	1.01e-07    &		2.69e-09	\\
0.5	&	4	&	2.48e-02 &	1.47e-04 &	1.30e-06    &		7.39e-08	\\
0.5	&	8	&	5.03e-02 &	4.91e-04 &	1.21e-05    &		2.17e-06	\\
1	&	2	&	7.22e-02 &	1.33e-05 &	3.18e-08    &		6.96e-10	\\
1	&	4	&	2.10e-02 &	1.03e-04 &	8.35e-07    &		5.50e-08	\\
1	&	8	&	4.65e-02 &	4.06e-04 &	9.42e-06    &		1.86e-06	\\
 \hline	
\end{tabular}
\caption{Iteration $1,2,3$ and $4$ of the algorithm, reduced model with dynamic rods scenario.}\label{tab:a2}
\end{table}
%--------------------------------------------------
\begin{figure}[!hp]
\begin{minipage}[c]{0.48\linewidth}
\centering\includegraphics[width=8cm,height=6cm]{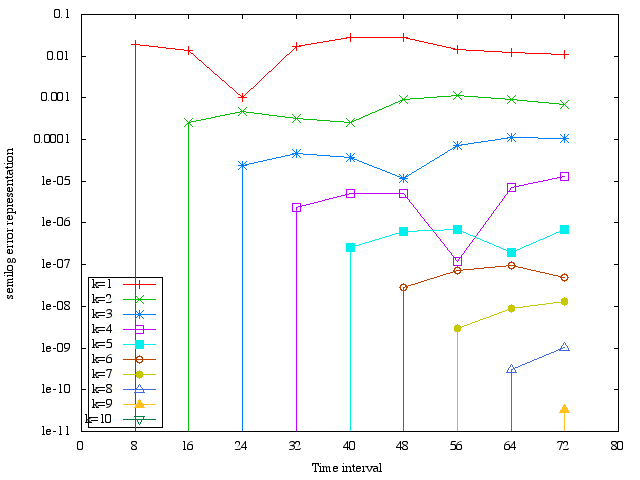}
\end{minipage}
\begin{minipage}[c]{0.48\linewidth}
\centering
\includegraphics[width=8cm,height=6.4cm]{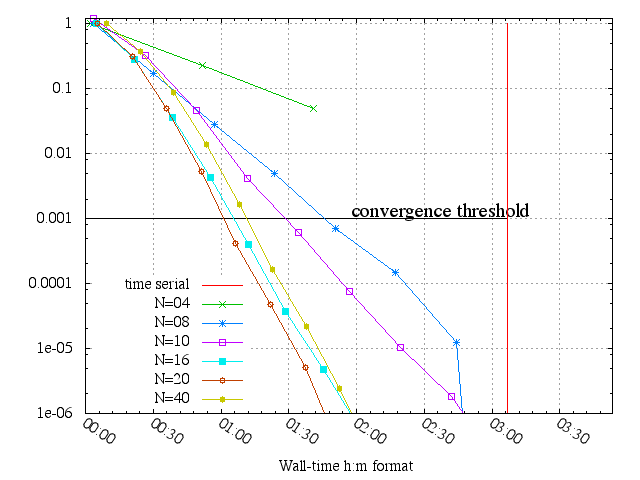}
\end{minipage}
\caption{ (left) Convergence of the algorithm (iteration 00-09) for the LWM transient problem coupled with the reduced model: $\tau_{F}=10^{-1}$, $\tau_{G}=4$ and $N =10$.
	 (right) Wall-time simulation (format: h:m) parareal in time algorithm and the serial one, we vary the number of used processor/subinterval; $\tau_{F}=10^{-1},\tau_{G}=2$ and $N \in\{1,4,8,10,16,20,40\}$.}
\label{n=var}
\end{figure}

%	We present in Figure~\ref{n=var} the acceleration of the resolution according to the number of subdivisions by temporal sub steps. The resolution on each sub steps is entrusted to a processor of the machine massively parallel. We note that passing from $8$ processors to $16$ processors divide approximatively by two the computational time. In fact, it makes it possible to pass from $2h:45$mn to $1h:15$mn of treatment. This is in adequacy with a local effectiveness of a machine with $4$ cores.\\ 
%Notes that if the number of the subdivisions $\hat n$ increases then the number of the processors agents in parallel also increases. The communication thus becomes increasingly important. This explains the fact that the curve of error in time wallclock (CPU) relating to the use of $40$ processors is worse than that for $20$ processors. The machine on which we submitted these numerical tests thus reached its saturation of speedup. 

	The basic parallel implementation of the parareal in time algorithm speeds up the resolution with a good efficiency using $16$ processor unit (see figure 7 left). Even it uses the DDM tools, the parareal in time algorithm is technically different. Indeed the communications are strongly damping the computational time since it occurs in one sense i.e. from time 0 to time T. Another point that contrast the parareal from the DDM is the package of information that should pass across the communicator. In fact, this package increases with the number of the decomposition contrary to the DDM. We refer to \citep{BLMM} for alternative parallel implementations.
  
\section{Conclusion}
We have present in this paper an application of the parareal algorithm in the paralallelization across the time of the simulation of the neutron diffusion multi-group kinetic equations. In order to improve computational time, the model is reduced by the use of an adequate corse solver based on several property of the steady solution. Numerical results shows that the algorithm speeds-up the simulation and converges quite fast, when the thresholds is reached in most case in two or three iterations of the parareal algorithm.   

\section*{Acknowledgement}
The authors gratefully acknowledge the support of LRC-MANON\footnote{Joint research program MANON between  CEA-Saclay and University Pierre et Marie Curie-Paris 6} %through grant {\color{red}Nxxxxxxxx}
, who also supported the first author during the preparation of a part of his PhD thesis~\citep[chap.3]{PhDriahi}.

%%%%%%%%%%%%%%%%%%%%%%%%%%%%%%%%%%%%%%%%%% ANNEX
\newpage
\section*{Appendix}

For the sake of convenience, we recall here the various physical constants used in the numerical simulation of the Langenbuch-Maurer-Werner (LMW) benchmark~\citep{LMWref}.
\begin{table}[!h]
\centering%\footnotesize
\begin{tabular}{ccccc} 
   \hline\hline
    Physical data & \multicolumn{2}{r}{medium} \\\hline\hline
 Cross sections & \multicolumn{2}{c}{fuel A} &\multicolumn{2}{c}{fuel B}\\
                 &  group-1 & group-2&     group-1& group-2 \\
\cline{2-5} 
$\Sigma_{t}$&0.23409670&0.93552546&0.23381787&0.95082160\\
$\Sigma_{f}$&0.006477691&0.1127328&0.007503284&0.1378004\vspace{.1in}\\  
 $celerity\, V^{g}$&1.25e+7&2.5e+5&1.25e+7&2.5e+5.\\\hline
& \multicolumn{2}{c}{rods} &\multicolumn{2}{c}{reflector}\\
                 &  group-1 & group-2&     group-1& group-2 \\
\cline{2-5} 
$\Sigma_{t}$&0.23409670&0.93552546&0.20397003&1.26261670 \\
$\Sigma_{f}$&0.006477691&0.1127328&.0&.0\\
 $celerity\, V^{g}$&1.25e+7&2.5e+5&1.25e+7&2.5e+5.\\\hline
\end{tabular}  \caption{Total and fission cross-sections of the LMW 3d benchmark.}
\label{tab:tdn0}
\end{table}

\begin{table}[!h]%\footnotesize
\centering\begin{minipage}[c]{0.45\linewidth}\hbox{\underline{\it\bf fuel A}}
\begin{tabular}{c||c|c}
$\Sigma_{s}^{g' \rightharpoonup g} $& group-1 & group-2 \\\hline\hline
 group-1 & .20613914 &.01755550  \\\hline
  group-2 & .0 &.84786329 
\end{tabular}
\end{minipage}
\begin{minipage}[c]{0.45\linewidth}\hbox{\underline{\it\bf fuel B}}
\begin{tabular}{c||c|c}
$\Sigma_{s}^{g' \rightharpoonup g}$ & group-1 & group-2 \\\hline\hline
 groupe-1 &.20564756  & .01717768 \\\hline
  groupe-2 &.85156526  &.0 
\end{tabular}
\end{minipage}\\
\begin{minipage}[c]{0.45\linewidth}\hbox{\underline{\it\bf rods of control}}
\begin{tabular}{c||c|c}
$\Sigma_{s}^{g' \rightharpoonup g} $& groupe-1 & groupe-2 \\\hline\hline
 groupe-1 &.20558914  &.01755550  \\\hline
  groupe-2 &.84406329  & .0
\end{tabular}
\end{minipage}
\begin{minipage}[c]{0.45\linewidth}\hbox{\underline{\it\bf reflector}}
\begin{tabular}{c||c|c}
$\Sigma_{s}^{g' \rightharpoonup g} $& groupe-1 & groupe-2 \\\hline\hline
 groupe-1 &.17371253  &.02759693  \\\hline
  groupe-2 &1.21325319  & .0
  \end{tabular}
\end{minipage}\caption{``scattering'' cross section data.}
\label{tab:tdn1}
\end{table}

\begin{table}\centering %\hbox{\underline{\it\bf Precursors}}
\begin{tabular}{ccccccc}\\\hline\hline
\text{precursor} & \text{group-1} & \text{group-2} & \text{group-3} & \text{group-4} & \text{group-5} & \text{group-6} \\ \hline\hline
$\lambda(s^{-1})$ &0.0127& 0.0317& 0.115& 0.311& 1.4 &3.87 \\\hline
$\beta$ &0.000247& 0.0013845& 0.001222& 0.0026455& 0.000832 &0.000169\\
\end{tabular}\caption{Precursors data}\label{precdata}
\end{table}

%%%%%%%%%%%%%%%%%%%%%%%%%%%%%%%%%%%%%%%%%%
\newpage

\bibliographystyle{elsarticle-num-names}\bibliography{ref_neutronique.bib}
%-----------------
%
%  DONT FORGET TO ADD THOSE REFERENCES
%
%------------------


 %\begin{thebibliography}{00}
%[Dahmani, 1999] Dahmani M., Baudron A.M., Lautard J.J. and Erradi L., Application of the nodal mixed dual technique of spatial reactor kinetics using the improved quasi static method, Proc. ANS Topical Mtg., Mathematics and Computation, Reactor Physics and Environmental Analysis in Nuclear Applications, Volume 1, Madrid, Spain, 1999.
%
%
% [Lautard, 1999] Lautard J.J., Schneider D. and Baudron A.M., Mixed Dual Methods for Neutronic Reactor Core Calculation in the CRONOS System, Proc. ANS Topical Mtg., Mathematics and Computation, Reactor Physics and Environmental Analysis in Nuclear Applications, Volume 1, Madrid, Spain, (1999).
% 
% 
%[Langenbuch, 1977] Langenbuch, S., Maurer, W., Werner, W., Coarse-Mesh Flux-Expansion Method for the Analysis of Space-Time Effects in Large Light Water Reactor Cores, Nucl. Sci. Eng., 63, 437, 1977.
%%% \bibitem must have the following form:
%%%   \bibitem{key}...
%%%
%
%% \bibitem{}
%
% \end{thebibliography}

\end{document}